\documentclass[aps, prb, twocolumn, superscriptaddress, floatfix]{revtex4-2}
\usepackage{graphicx}
\usepackage{amsmath}
\usepackage{amssymb}
\usepackage{tikz}
\usepackage{braket}
\usepackage{pgfplots}
\usepackage[version=4]{mhchem}
\pgfplotsset{compat=1.17}
\usepackage{subcaption}
\usepackage{enumitem}
\usepackage{mathrsfs}
\usetikzlibrary{shapes.geometric}
\usetikzlibrary{shadows,patterns,perspective}
\usetikzlibrary {decorations,decorations.text}
\usetikzlibrary{decorations.pathreplacing}
\usepackage{float}
\usepackage{hyperref}
\usepackage{xcolor}
        
\definecolor{linkcolor}{RGB}{0, 0, 255}      
\definecolor{citecolor}{RGB}{0, 128, 0}     
\definecolor{urlcolor}{RGB}{255, 0, 0}       

\hypersetup{
    colorlinks=true,
    linkcolor=linkcolor,
    citecolor=citecolor,
    urlcolor=urlcolor,
    linktoc=all,  
    pdfborder={0 0 0}  
}

\DeclareCaptionJustification{justified}{\leftskip=0pt \rightskip=0pt \parfillskip=0pt plus 1fil}

\begin{document}
\definecolor{dy}{rgb}{0.9,0.9,0.4}
\definecolor{dr}{rgb}{0.95,0.65,0.55}
\definecolor{db}{rgb}{0.5,0.8,0.9}
\definecolor{dg}{rgb}{0.2,0.9,0.6}
\definecolor{BrickRed}{rgb}{0.8,0.3,0.3}
\definecolor{Navy}{rgb}{0.2,0.2,0.6}
\definecolor{DarkGreen}{rgb}{0.1,0.4,0.1}

\title{Dissipation driven phase transition in the non-Hermitian Kondo model}
\author{Pradip Kattel}
\email{pradip.kattel@rutgers.edu}
\affiliation{Department of Physics, Center for Material Theory, Rutgers University,
Piscataway, NJ 08854, United States of America}

\author{Abay Zhakenov}
\affiliation{Department of Physics, Center for Material Theory, Rutgers University,
Piscataway, NJ 08854, United States of America} 

\author{Parameshwar R. Pasnoori}
\affiliation{Department of Physics, University of Maryland, College Park, MD 20742, United
States of America}
\affiliation{Laboratory for Physical Sciences, 8050 Greenmead Dr, College Park, MD 20740,
United States of America}

\author{Patrick Azaria}
\affiliation{Laboratoire de Physique Th\'orique de la Mati\`ere Condens\'ee, Sorbonne Universit\'e and CNRS, 4 Place Jussieu, 75252 Paris, France}

\author{Natan Andrei}
\affiliation{Department of Physics, Center for Material Theory, Rutgers University,
Piscataway, NJ 08854, United States of America}

\begin{abstract}
Non-Hermitian Hamiltonians, as effective models, capture phenomena such as energy dissipation and non-unitary evolution in open quantum systems. New phases and phenomena appear that are not present in their Hermitian counterparts. Such a Hamiltonian, the non-Hermitian Kondo model, {  has been used to describe inelastic scattering between mobile and confined atoms} in an optical lattice \cite{nakagawa2018non}. Using a combination of Bethe Ansatz and perturbative calculation, the authors argued that this model has two distinct phases: the Kondo phase and the non-Kondo phase, where impurity is screened and unscreened, respectively.  We show, however, that a novel phase termed $\widetilde{YSR}$ emerges between the Kondo and unscreened phases.
Characterized by two RG invariants: a generalized Kondo temperature ($T_K$) and a loss strength parameter ($\alpha$), the system exhibits three distinct phases. In the increasing order of losses, they are: The Kondo phase ($0<\alpha<\pi/2$), the $\widetilde{YSR}$ phase ($\pi/2<\alpha<3\pi/2$), and the local moment phase ($\alpha>3\pi/2$). Notably, phase transition driven by dissipation occurs across $\alpha=\pi/2$, where both energetics and different time scales associated with loss play roles.

\end{abstract}

\maketitle
\section{Introduction}
\begin{figure}
\begin{center}
\definecolor{dy}{rgb}{0.9,0.9,0.4}
\definecolor{dr}{rgb}{0.95,0.65,0.55}
\definecolor{db}{rgb}{0.5,0.8,0.9}
\definecolor{dg}{rgb}{0.2,0.9,0.6}
\begin{tikzpicture}[x=0.75cm, y=0.35cm, every node/.style={scale=1.}]
\draw [<->, rounded corners, thick, purple] (-1,2) -- (-1,-1.75) -- (2,-1.75);

\node at (-1,2.45) {$\Re$(E)};
\node at (2.2,-1.75) {{$\alpha$}};

\draw [fill=dr] (6,0)--(8,0)--(8,8)--(6,8);
\draw [fill=db] (2,0)--(6,0)--(6,8)--(2,8);
\draw [fill=dg] (0,0)--(2,0)--(2,8)--(0,8)--(0,0);
\draw[dashed] (4,0)--(4,8);
\draw[blue,thick] (2,4)--(8,4);
\draw[red,thick] (2,2.5) .. controls (2.5,2.4) and (3.5,3.5) .. (4,4);
\draw[red,thick] (4,4) .. controls (5.5,5.5) and (5.5,5.4) .. (6,5.5);

\node at (0,-.75) {{0}};
\node at (2,-.75) {{${\pi}/{2}$}};
\node at (4,-.75) {{$\pi$}};
\node at (6,-.75) {{$3/2 \pi$}};

\node at (1,6) {\large{K}};
\node at (4.1,6.5) {\large{$\widetilde{YSR}$}};
\node at (7,6) {\large{LM}};



\node at (1, 3) {\large{$\ket{K}$}};
\node at (7, 3) {\large{$\ket{U}$}};
\node at (3, 3.0) {\large{$\ket{B}$}};
\node at (3, 4.5) {\large{$\ket{U}$}};
\node at (5, 3.35) {\large{$\ket{U}$}};
\node at (5, 4.76) {\large{$\ket{B}$}};




\end{tikzpicture}
\caption{{ {Phase diagram as a function of $\alpha$, which controls deviation from Hermiticity. In the Kondo phase (K), the impurity is fully screened in $\ket{K}$, while in the LM phase, it is partially screened in $\ket{U}$. In the $\widetilde{YSR}$ phase, the impurity is screened by a bound mode in $\ket{B}$ but only partially in $\ket{U}$. The red curve shows $\Re(E)$ of the bound mode, and the blue curve shows $\Re(E)$ of solutions on the complex locus ${\cal C}$ from Eq.\eqref{imagpart}, representing an unscreened state. Here, $\alpha$ is an RG invariant parameter (ref Eq.\eqref{alpha-def}).}}}
\label{Figure1}
\end{center}
\end{figure}
Dissipation is ubiquitous, even in well-engineered quantum platforms, necessitating careful study of its effects on any phenomenon being
described theoretically or measured in experiment\cite{pepino2010open,goldman2016topological,goldman2014light,lewenstein2006travelling,myerson2022construction,tsomokos2010using}. Dissipative phenomena are often effectively described using non-Hermitian Hamiltonians, which represent an open quantum system coupled to a large environment, allowing for the exchange of energy or particles. Such a set-up is most often described by a Lindbaldian  formulation \cite{gorini1976completely,lindblad1976generators}
or by a Feshbach projection approach \cite{ashida2020non},  which, under appropriate conditions, can be reduced to a non-Hermitian Hamiltonian \cite{nakagawa2018non,han2023complex} when the environment is integrated out. Such non-Hermitian Hamiltonians incorporate dissipative effects and lead to many effects without Hermitian counterparts \cite{zhang2021observation,zou2021observation,zhang2021acoustic,wang2021observation,sakhdari2019experimental,schnabel2017pt,shi2016accessing}.

\section{Model}

Here, we study the Kondo system in dissipative media \cite{riegger2018localized,kanasz2018exploring,white2020observation}, revealing novel effects beyond the standard impurity screening.
One such effect where the impurity remains unscreened in the local moment regime was noted in \cite{nakagawa2018non}; In this work we reveal a regime where the impurity is screened by a single particle bound mode, among other novel effects. As  shown in~\cite{nakagawa2018non}, the non-Hermitian Hamiltonian, 
\begin{equation}
H=-i \int_{0}^L \psi^{\dagger}(x) \partial_x \psi(x)\;  d x+J\;   \psi^{\dagger}(0)\vec \sigma \psi(0) \cdot \vec S 
\label{Hamiltonian}
\end{equation}
describes the Kondo effect in a dissipative AMO system consisting of two-orbital $^{173}$Yb gas atoms where the atoms in the metastable excited state play the role of spin S = 1/2 impurities. Here $\psi^T(x) \equiv (\psi_{\uparrow}(x), \psi_{\downarrow}(x))$ is the { two-component} spinor field describing the itinerant 
atoms, $\vec \sigma$ are the Pauli matrices and $\vec S$ denotes the impurity spin localized at $x=0$.
In Eq.(\ref{Hamiltonian}), the interaction coupling  $J=J_r+iJ_i$  is complex-valued 
 and its imaginary part is related via the electron density $D=N^e/L$ to the rate of two-body losses due to inelastic scattering  $|\Gamma_0|= DJ_i$.  The interplay of the two-body losses and the Kondo effect leads to new dynamical phenomena and a new phase transition.
 Nakagawa et al showed
 that the Kondo effect survives a small imaginary Kondo coupling $J_i\neq 0$, but for large enough $J_i$, the impurity is unscreened. Here we shall show that on top of these two phases, there exists a new phase,  coined Yu-Shiba-Rushinov($YSR$)-like ($\widetilde{YSR}$) \footnote{The YSR phase was first discovered in a BCS superconductor by Yu, Shiba and Rusinov who pointed out the existence of these bound modes}, where the impurity can be screened by a {\it single particle bound mode}. { As this bound mode is characterized by a finite energy scale, there is no fixed point associated with it,  unlike in the other phases. Therefore, the renormalization group approach of \cite{nakagawa2018non} could not identify this phase.

The appearance of this intermediate phase is quite universal; it appears in wide ranges of Hermitian models like spin chain with magnetic impurities \cite{kattel2023kondo}, superconductor with Kondo impurity \cite{Kondosup}, and $\mathcal{PT}-$ symmetric non-Hermitian Kondo impurity in spin chain \cite{kattel2024spin}. }

\smallskip

Like its Hermitian counterpart, the Hamiltonian Eq.(\ref{Hamiltonian}) is integrable \cite{nakagawa2018non} with its Bethe Ansatz equations being
the analytical continuation of those of the Hermitian case \cite{andrei1992integrable,andrei1983solution, tsvelick1983exact} to complex coupling. The energy eigenvalues are then complex ${\cal E}= E +i \Gamma(E)$ as we show below,
with the imaginary part determining the lifetime decay or enhancement of the state depending on whether $\Gamma(E)$ is negative or positive. Being complex, there is no natural ordering of the spectrum, and the relevance of a given state depends on both $E$ and $\Gamma(E)$, as we shall see below. 
  Conventionally,  the eigenstate with the minimal real part of the (complex) energy is termed the ground state. However, as the imaginary parts of the energies affect the nonunitary evolution of the system, the amplitudes of states with positive (negative) imaginary parts may be enhanced (suppressed) during time evolution, irrespective of the real part of the energy. This reflects the interplay between minimizing the energy and the dynamical stability in lossy non-Hermitian systems.

The various phases in the model are characterized by two renormalization group invariants:  $T_K $, the Hermitian Kondo temperature, and  $\alpha$, a measure of the departure from Hermiticity, related in the scaling limit to the bare Kondo couplings by $( 2\alpha /\pi)  \simeq J_i/ J_r^2$ for $J_i  \ll 1   $. More precisely, we show, 
\begin{equation}
T_K= 2D e^{-\frac{\pi \cos \phi}{c}}, \; \alpha=\pi \sin \phi / c\;,
\label{alpha-def}
\end{equation}
where $c$ and $\phi$ are related to the complex coupling constant $J $ as follows: $\frac{2 J}{1-\frac{3 J^2}{4}} = c \; e^{i \phi}, \; c \in \mathbb{R}$. Both $T_K$ and $\alpha$ are held fixed in the scaling limit $D\to \infty, c \to 0$ where the results are universal.

The new $\widetilde{YSR}$ phase found in this work lies in the regime $\pi/2\le \alpha < 3 \pi/2$ where in addition to the usual solutions of the Bethe equations a new solution appears, called impurity string (IS) solution. The real part of its energy  is given by
$E_b=-T_K\sin \alpha$. It changes sign at $\alpha=\pi$, { {so that occupying the IS in the range $\pi/2 \le \alpha \le \pi$ lowers the real part of the energy leading to the state $\ket{B}$, while in the range $\pi \le \alpha \le 3\pi/2$, not occupying the IS lowers the energy leading to a state $\ket{U}$}}. In the state $\ket{B}$, the impurity is screened by a bound mode localized near it.  


The bound mode energy also has an imaginary part $\Gamma_b = T_K \cos \alpha <0$ which gives the bound mode a finite lifetime. Hence, in the $\widetilde{YSR}$ phase, in the regime $\pi/2\le \alpha < \pi$,  the impurity is eventually found to be unscreened at long enough times $\Gamma_0^{-1} \gg  t \gg  |\Gamma_b|^{-1}$. This highlights that the quantum phase transition from the Kondo to local moment phases is dynamically induced by losses.

\smallskip

\section{Bethe Ansatz equations}
 We now turn to the derivation of these results from the Bethe Ansatz. The spectrum of the Hamiltonian  is given by (see appendix A),
\begin{equation} 
E=\sum_j k_j= \sum_{j=1}^{N^e} \frac{2\pi}{L}n_j + D \sum_{\gamma=1}^{M} \log    \frac{\Lambda_{\gamma}-1+\frac{ic'}{2}}{\Lambda_{\gamma}-1-\frac{ic'}{2}}\;. 
\label{engeng}
\end{equation}
Here $N^e$ denotes the number of electrons,
$D=N^e/L$ is their density, the integers $n_j$ are the charge quantum numbers, and $c'=\frac{2 J}{1-\frac{3 J^2}{4}} = c \; e^{i \phi}, \; c \in \mathbb{R}$.   The spin rapidities $\Lambda_{\gamma}, \gamma=1 \dots M$ govern the spin dynamics and satisfy,

\begin{align}
    \sum_{\gamma=1}^{N^e+1}\Theta(2(\Lambda_\gamma-1+\mu_\gamma)) =-2\pi I_\gamma+\sum_{\delta=1}^M\Theta(\Lambda_\gamma-\Lambda_\delta)\;,
    \label{betheequations}
\end{align}
where $\Theta(x)=-2\arctan(x/c)$, $\mu_\gamma$ is 0 for electrons and $e^{-i\phi}$ for impurity and $I_{\gamma}$ are integers (half integers) depending on $N^e-M$ being even (odd) whose choice specifies the state \footnote{ Our Bethe equations are related from those of \cite{nakagawa2018non} by a change of variables $\Lambda_{\gamma}\to 1+\Lambda_{\gamma} c$}. In the thermodynamic 
limit, the solutions of (\ref{betheequations}) form a dense set which lies on a curve ${\cal  C}$ in the complex plane{ : $\Lambda=\mu+i\nu(\mu)$ with 
\begin{equation}
\nu(\mu)=\frac{c \cosh(\pi(\mu-1)/c)}{ 2\pi N^e}\log \frac{\cosh(\frac{\pi(\mu-1+\cos\phi)}{c})+\sin(\alpha)}{\cosh(\frac{\pi(\mu-1+\cos\phi)}{c})-\sin(\alpha)}\;.
\label{imagpart}
\end{equation}
Note that the imaginary part $\nu(\mu)$ is of order $O(1/N^e)$. The roots $\Lambda$ of the Bethe equation are dense in the complex plane, as shown in Fig.`\ref{fig:SBAE-loci}.
\begin{figure}
    \centering
\includegraphics[width=0.9\linewidth,height=3.5cm]{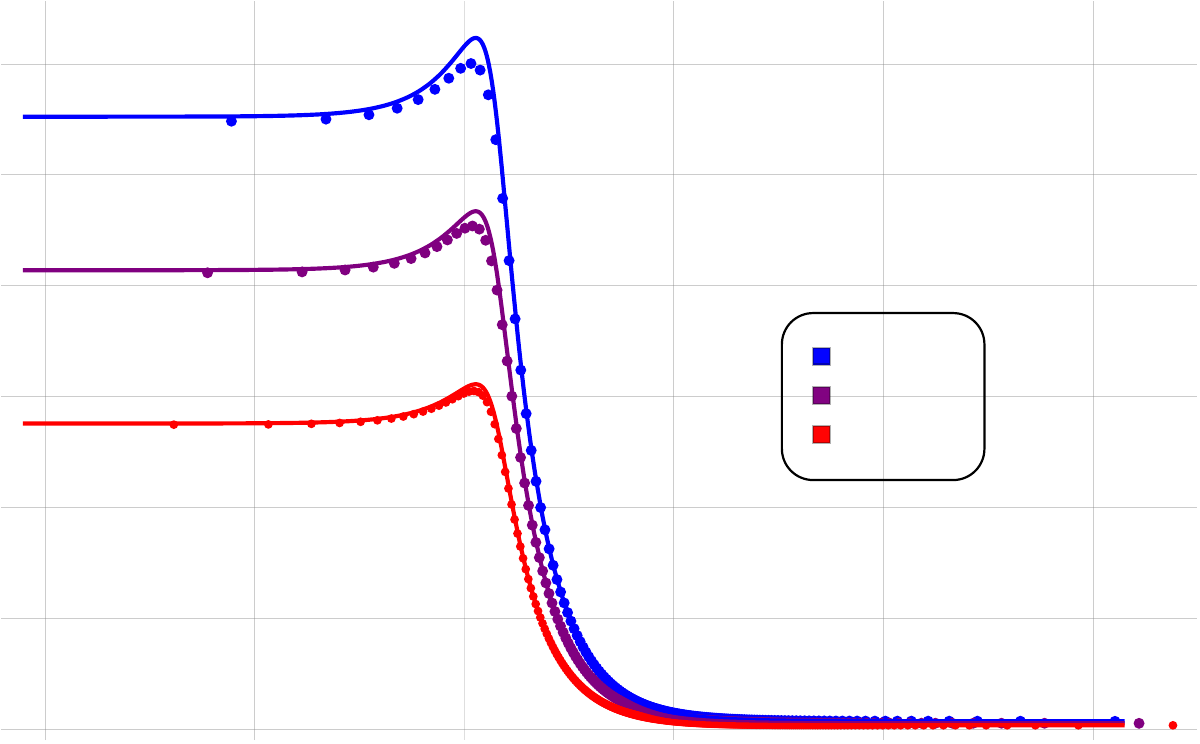}
    \begin{tikzpicture}[overlay, remember picture]
    \node at (-2,1.85) {\tiny{$N=300$}};
    \node at (-2,1.65) {\tiny{$N=400$}};
     \node at (-2,1.45) {\tiny{$N=600$}};
      \node at (-3.55,-0.1) {$1$};
     \node at (-2.15,-0.10) {$2$};
      \node at (-0.65,-0.10) {$3$};
      \node at (-4.85,-0.10) {$0$};
      \node at (-6.35,-0.10) {$-1$};
      \node at (-7.75,-0.1) {$-2$};
      \node at (0,-0.15) {$\Re(\Lambda)$};
      \node at (-8.25,1.75) {\rotatebox{90}{$\Im(\Lambda)$}};
      \node at (-8,0.2) {$0$};
      \node at (-8,1.175) {$0.004$};
      \node at (-8,2.2) {$0.008$};
      \node at (-8,3.2) {$0.012$};
    \end{tikzpicture}
    \caption{Loci of Eq.\eqref{betheequations} in the complex plane for $c=1.25$ and $\phi=0.54$. Dotted lines are numerical BAE solutions, and solid lines are exact results given by Eq.\eqref{imagpart}.}
    \label{fig:SBAE-loci}
\end{figure}
}

The density of solutions $\sigma(\Lambda) $  on the curve  ${\cal  C}$ can be obtained  from
Eq.\eqref{betheequations}  in thermodynamic limit as \cite{andrei1992integrable}
\begin{equation}
\sigma(\Lambda)=f(\Lambda)-\int_{{\cal  C}}\mathrm d\Lambda'\;   K(\Lambda-\Lambda')\sigma(\Lambda')\;,
\label{sdens}
\end{equation}
where the kernel is given by $K(\Lambda)=\frac{1}{\pi}\frac{c}{c^2+\Lambda^2}$ and  the function $f(\Lambda)$  depends on
the state. The number of roots in dense set ${\cal  C}$ is  $ M=
\int_{{\cal  C}}\mathrm d\Lambda \; \sigma(\Lambda) $ and   the  spin of the state is $
S=\frac{N^e+1}{2} - \int_ {\cal  C}\mathrm d\Lambda\;\sigma(\Lambda).$



On top of the dense set of roots in $ {\cal  C}$,
there exists  in  the limit $N^e \rightarrow \infty$, for  $\pi/2 <\alpha < 3 \pi/2$, an additional  {\it isolated} solution
 of  Eq.(\ref{betheequations})  
\begin{equation}
    \Lambda_{IS} =1-\frac{i c}{2}-e^{-i \phi }\;.
        \label{IS}
\end{equation}
This solution, the Impurity String (IS),  \footnote{Such a complex solution of Bethe Ansatz describing a dissipative mode was first found in \cite{kattel2023exact}.}
describes a bound mode in the regime $\pi/2 <\alpha < 3 \pi/2$ which may or not be occupied. As we show below, it is responsible for the new $\widetilde{YSR}$ phase.


\subsection{The Kondo phase}

We proceed to discuss the various phases of the model. The Kondo phase, when $0\le \alpha < \pi/2$,  the state $\ket{K}$ is obtained by choosing consecutive quantum numbers $I_{\gamma}$ leading to the density,
\begin{equation}
\sigma_K(\Lambda)= \frac{N^e}{2c}\mathrm{sech}\left( \frac{\pi}{c}(\Lambda-1)\right)  +\frac{1}{2c}\mathrm{sech}\left( \frac{\pi}{c}(\Lambda-1+e^{-i\phi})\right)
\label{sigmakondo}
\end{equation}
from which we get for the total number of roots $M=\int_{\cal C} \mathrm d\Lambda\; \sigma_K(\Lambda) =\frac{N^e+1}{2}$  which requires
$N^e$ to be odd. From $S=M-\frac{N^e+1}{2}$, we find that $S=0$ and hence that 
the impurity is screened.
The ground state energy is 
\begin{equation}
E_{0K}=-\frac{\pi N^2}{2L} -i D\log{\left( \frac{\Gamma(\frac{1}{2} -\frac{i}{2c'} )\Gamma(1+\frac{i}{2c'})}{\Gamma(\frac{1}{2} +\frac{i}{2c'})\Gamma(1-\frac{i}{2c'})} \right)},
\label{gsenergykondo}
\end{equation}
 with both real and imaginary components when $\phi \neq 0$. The real part $\Re{(E_{0K})}$ is the energy of the state, and the imaginary part $\Im{(E_{0K}})$ corresponds to the inverse lifetime of the Kondo state. It is given in the scaling limit to leading order in the asymptotic expansion
of Eq.(\ref{gsenergykondo}), by $\Im(E_{0K})= -\Gamma_0+ {\cal O}(D/\log(2D/T_K))$ which is the bare decay rate of the two body losses.

\begin{figure*}[ht!]
    \begin{minipage}{0.30\textwidth}      \includegraphics[width=\linewidth]{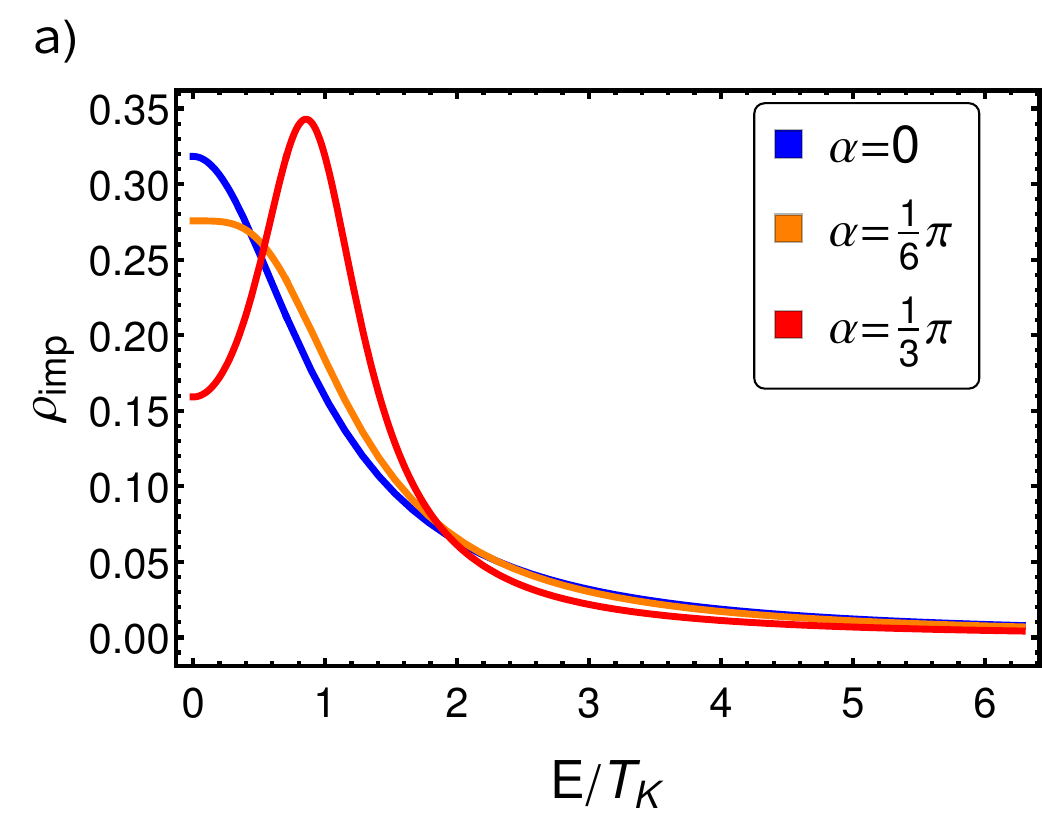}
    \end{minipage}%
    \begin{minipage}{0.30\textwidth}
        \includegraphics[width=\linewidth]{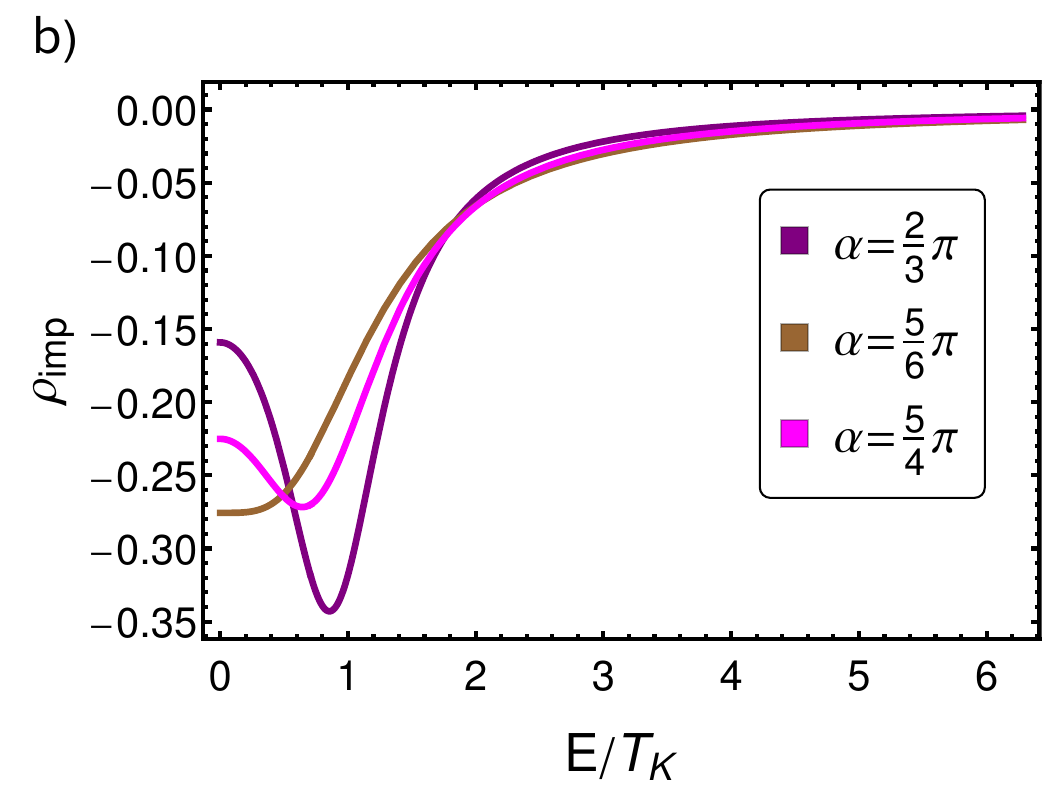}
    \end{minipage}%
    \begin{minipage}{0.30\textwidth}
        \includegraphics[width=\linewidth]{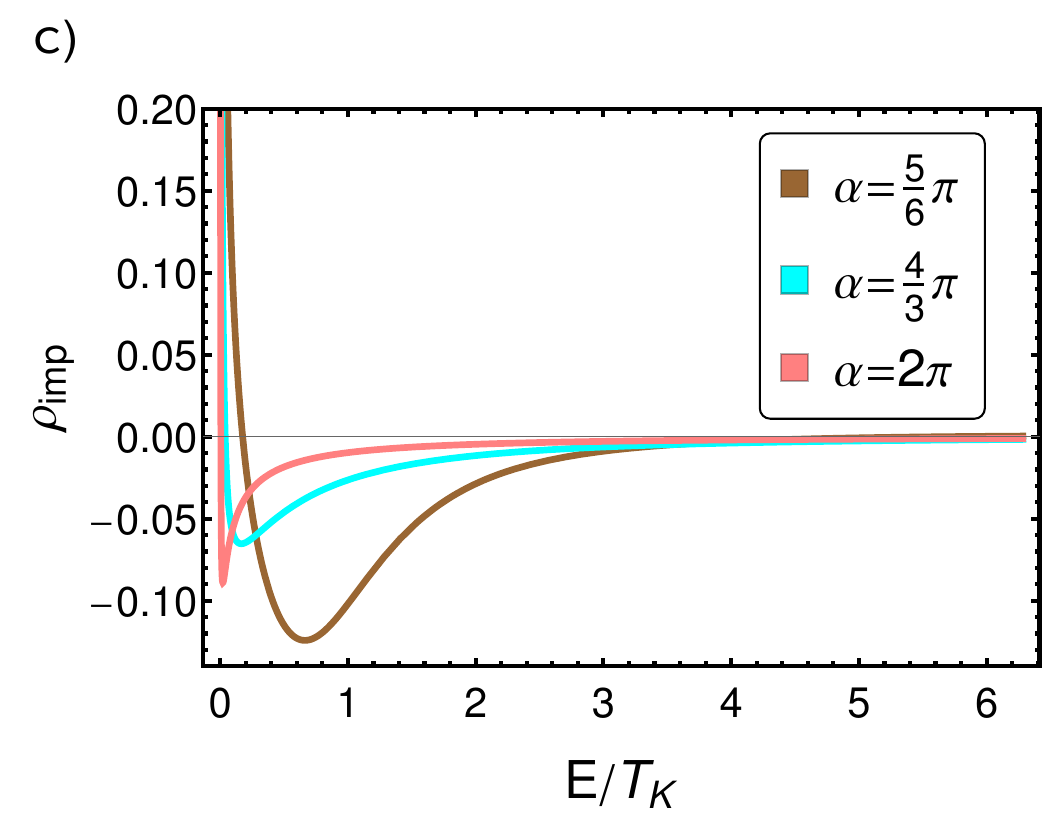}
    \end{minipage}
    \caption{Impurity contribution   to the one particle density of states $\rho_{imp}$ as a function of  real part of the energy $E$. a) $\rho_{imp}$ in the Kondo phase is shown for the Hermitian case when $\alpha=0$  (blue),  for $\alpha=\pi/6$   (orange) and for $\alpha =\pi/3$ (red). We notice that the maximum of $\rho_{imp}$ starts to shift smoothly from $E=0$ to $E=T_K$ as $\alpha$ varies from $\pi/6$ to $\pi/2$.  In b) we show the continuous root distribution contribution to $\rho_{imp}$ in the $\ket{B}$ state in the $\widetilde{YSR}$ phase. The complete $\rho_{imp}$ includes also a delta function contribution $\delta(E-E_b)$, not shown in the figure,  from the isolated impurity string Eq.\eqref{dosdelta}.  c) $\rho_{imp}$ in $\ket{U}$ state for various values of $\alpha$ in the $\widetilde{YSR}$ and LM phase.}
    \label{DOSfigs}
\end{figure*}

The simplest excitations above the ground state are constructed by creating ``holes" in the ground state sequence of $I_{\gamma}$. These excitations, the spinons,  carry spin $ \frac{1}{2}$ \cite{andrei1980diagonalization}
and have complex energies relative to the ground state (in the scaling limit  $|{\cal E}| \ll D$),
\begin{equation}
{\cal E}=2 D \displaystyle{e^{\frac{\pi}{c} (\Lambda-1)}}\;,
\label{spinon}
\end{equation}
where $\Lambda $ is the hole position {lying along the complex curve ${\cal  C}$ described above \eqref{imagpart}. The spinon energies form a complex curve $\mathcal{E}=E+i\Gamma(E)$ with the imaginary part given by
\begin{equation}\label{imgenergy}
\Gamma(E)=\frac{1}{L}\;  \tanh^{-1} \left(\frac{2ET_K}{E^2+T_K^2} \sin \alpha\right)\;.
\end{equation}
The corresponding complex-valued density of states is obtained using
$\rho({\cal E}) = \sigma(\Lambda) \frac{d \Lambda}{d{\cal E}}$ along the complex curve ${\cal  C}$ yielding
\begin{equation}
\rho_K({\cal E}) =\frac{L}{2 \pi}+ \frac{1}{\pi} \frac{T_0}{{\cal E}^2+T_0^2}\;,
\label{kondodos}
\end{equation}
where $T_0=T_K e^{i\alpha}$. The first term in Eq.(\ref{kondodos}) is the DOS of the bulk fermions, whereas
the second term is the contribution of the impurity, which displays the characteristic Lorenzian shape of the Hermitian Kondo problem. 
Using $\Gamma(E)$ from \eqref{imgenergy}, one can obtain the real-valued density of states $\tilde \rho_K(E)$ as follows
\begin{equation}
\tilde \rho_K(E) d E =\rho_K(E + i \Gamma(E))\times(1+i\partial \Gamma(E)/\partial E) d E \in  \mathbb{R}
\label{rhotilde}
\end{equation}
leading to
\begin{eqnarray}
\tilde \rho_K(E) &=& \Re(\rho_K(E)) + {\cal O}(1/L)\;. 
\label{rrrhotilde}
\end{eqnarray}
Here $\Re(\rho_K(E)) $ is the real part of the complex-valued density of states Eq.(\ref{kondodos}) for real energies.}

{  In the Kondo regime $\Gamma(E) > 0$}, hence the states with one spinon have a {\it larger}
 lifetime than the Kondo ground state itself. This indicates that it is dynamically advantageous 
 to remove a state from the  Kondo cloud as this lowers the amplitude for a singlet state to be formed at the impurity 
site, hence avoiding the possibility of losses.  However,  the time scale for such a process being $\propto L$, we expect the Kondo state
to be dynamically stable against depopulation of the screening cloud through single spinon excitations. 

Returning to the DOS of single-spinon excitations, we plot in Fig.(\ref{DOSfigs})(a) the contribution of the impurity $\rho_{\rm imp}= \tilde \rho_K(E)-L/2\pi$ to  the  DOS (Eq.(\ref{rrrhotilde})), i.e: 
\begin{equation}
\rho_{\rm imp}\left[\frac{E}{T_K}\right] =\frac{\cos \alpha}{\pi T_K} \frac{1+(E/T_K)^2}{1+ 2(E/T_K)^2 \cos 2\alpha + (E/T_K)^4}
\label{rhoK}
\end{equation}
as a function of the real part of the energy $E$. 
As $\alpha$ varies from $0$ to $\pi/2$, the impurity DOS changes from a pure Kondo behavior at $\alpha=0$ with a peak at $E=0$, to
a situation where the peak is shifted to a non-zero energy $E_{\alpha}=\sqrt{2 \sin \alpha -1}\;  T_K$ when $\alpha \ge \pi/6$. Such a shift may be observed by STM measurements \cite{pcref}.
Eventually, when $\alpha \rightarrow (\pi/2)^-$ it develops a delta peak at $E=T_K$: $\rho_{\rm imp} \rightarrow 1/2 \; \delta(E-T_K)$. 
As $\alpha$ increases (so does the bare loss rate $J_i$)
the number of modes that contribute to the screening of the impurity decreases until there remains one single mode at $E=T_K$ when $\alpha \rightarrow \pi/2$. We interpret this transfer of spectral weight towards $T_K$ as a prelude to the appearance of a bound mode when $\alpha > \pi/2$.

\subsection{The bound-mode phase}
We now consider the regime $\pi/2 < \alpha < 3\pi/2.$
  In this phase, the impurity string $IS$, Eq.(\ref{IS}), is a solution of the Bethe equations in the thermodynamic limit. Its energy  is given by (see appendix A) $E_{IS, { \rm charge}}=-\frac{\pi}{2L},\;\;\;\; E_{IS, spin}= -i T_0$, or setting $E_{IS, { \rm spin}}= E_b+ i \Gamma_b$, we have
$E_b=-T_K \sin \alpha \;  \;  \;  {\rm and}\; \;  \;  \Gamma_b= T_K \cos \alpha \le 0$.
The imaginary part of the bound mode energy $\Gamma_b$ is negative for any $\alpha$, indicating that the bound mode is dynamically unstable and has a \textit{finite} lifetime $ {\Gamma_b}^{-1}$. 

One obtains then two possible states, $\ket{B}$ and $\ket{U}$, depending on whether or not one adds the IS  to the dense set of solutions ${\cal C}$.  As remarked above, the state $\ket{B}$ has a lower real part of energy in the range $\pi/2\le \alpha\le \pi$ while  $\ket{U}$ it has a lower real part of energy when  $\pi\le \alpha\le 3\pi/2$. 
  
In the $\ket{B}$ state, we find that the continuous root density $\sigma_B(\Lambda)$ and ground state energy $E_{0B}$ are analytic continuation of those in the Kondo phase Eq.(\ref{sigmakondo})  to the region $\pi/2 < \alpha < 3\pi/2$, i.e:
\begin{equation}
\sigma_B(\Lambda) \equiv \sigma_K(\Lambda), \quad E_{0K}\equiv E_{0B}, \;   \; \alpha \in \left[\frac{\pi}{2}, \frac{3\pi}{2}\right]\;.
\label{densityB}
\end{equation}
In this regime, the total number of roots, including the impurity string, is given by
$M= \int_{\cal C} \mathrm d\Lambda\; \sigma_B(\Lambda)+1= \frac{N^e+1}{2}$ requiring $N^e$ to be odd  \footnote{We consider the $\ket{U}$ and $\ket{B}$ states  only in even $N^e$  and odd $N^e$ parity sector respectively. The other sectors require adding a hole, which may be sent to $-\infty$ without changing the results. }. Hence,
the total spin is $S=0$ as in the Kondo phase, and 
the impurity is screened in the $\ket{B}$ ground state.


We follow the same procedure as for the $\ket{K}$ state to compute the spinon DOS as a function of the real part of the energy ($E$) with respect to the $\ket{B}$ and $\ket{U}$ ground states. For the $\ket{B}$ state, we find 
\begin{equation}
  \tilde  \rho_B(E)=\Re \tilde{\rho}_K(E)+\delta(E-E_b)\;,
  \label{dosdelta}
\end{equation}
where the spinon contribution $\Re \tilde{\rho}_K(E)$ is the analytic continuation of the Eq.\eqref{rrrhotilde} to the regime $\alpha\in \left[\frac{\pi}{2},\pi\right]$ and the delta function contribution is due to the bound mode. 
Here and also in the  $\ket{U}$ ground states, we find $\Gamma(E)\sim \frac{1}{L}$ and positive. The impurity contribution to the spinon  DOS is shown in Fig.\ref{DOSfigs}(b) and (c), respectively, for the $\ket{B}$ and $\ket{U}$ states. Notice that the impurity contribution to the spinon DOS is always negative in the $\ket{B}$ state. However, due to  the positive delta function contribution 
in Eq.(\ref{dosdelta}), the integrated density of the state is positive and equals $1/2$ as in the Kondo phase.
We interpret the negative contribution of the spinons as the signature that spinons do not participate in the screening of the impurity in the $\ket{B}$  state. This negative DOS also was seen in Hermitian models with boundary-bound modes \cite{kattel2023kondo,kattel2024kondo}.
In the $\ket{U}$ state, we observe that the contribution of the impurity to the DOS can be positive as well as negative depending on the energy. Although it is not completely clear to us, we conjecture that the positive contribution corresponds to a partial screening of the impurity.

Turning to the state $\ket{U}$ which includes only the solutions in ${\cal C}$ we find  that the density $\sigma_U(\Lambda)$  is given by
\begin{equation}
\sigma_{U}(\Lambda)=\frac{1}{2 c}\frac{N^e}{\cosh \frac{\pi}{c}(\Lambda-1)}+ \frac{1}{2\pi} \left( \frac{1}{z} + \Psi\left(\frac{z}{2c}\right)-  \Psi\left(\frac{z+c}{2c}\right) \right)\;,
\label{densityU}
\end{equation}
where $z=i(\Lambda - \Lambda_{IS})$ and $\Psi(z)$ is the digamma function. The total number of roots is $M=\int_{\mathbb{R}} \mathrm d\Lambda\; \sigma_U(\Lambda) =\frac{N^e}{2}$ which requires $N^{e}$ to be even. This state has total spin $S^z=\frac{1}{2}$, which indicates that the impurity is unscreened. The total ground-state energy is given by
\begin{equation}
E_{0U}=-\frac{\pi N^2}{2L} -i D\log{\left( \frac{\Gamma\left(\frac{i}{2c'} \right)\Gamma\left(1+\frac{i}{2c'}\right)}{\Gamma^2\left(\frac{1}{2} +\frac{i}{2c'}\right)} \right)}\;.
\label{gsenergyU}
\end{equation}
The spinon DOS in this regime is given by (see appendix A)
\begin{equation}
    \tilde\rho_U(E)=\Re \rho_{U}(E)\;,
    \label{DOSU}
\end{equation}
where $\rho_U({\cal E})=\sigma_U(\Lambda)\frac{d\Lambda}{d{\cal E}}$. 

\subsection{The local moment phase}
Considering now the phase  $\alpha> 3\pi/2$, there is no single particle bound mode, and hence 
the impurity cannot be \textit{completely} screened. For $N^e$ even, the total spin
of the ground state is $S=1/2$, and it is described by the analytical continuation of the $\ket{U}$ to values of $\alpha>3\pi/2$. The ground state root distribution and the density of states are given by Eq.\eqref{densityU} and Eq.\eqref{DOSU}, respectively. The DOS as shown in Fig.\ref{DOSfigs} has positive and negative parts. We interpret that the impurity is partially screened by the positive part.

 
\section{Conclusion and outlook}
We conclude then that the Kondo system in an open quantum setting has a novel dynamical phase transition at $\alpha=\frac{\pi}{2}$. For $\alpha<\frac{\pi}{2}$, the Kondo physics survives with the impurity screened by the Kondo cloud. However, a new dynamical phase appears when $\frac{\pi}{2}<\alpha<\frac{3\pi}{2}$. Two distinct kinds of states appear, one where the impurity is screened by a bound mode and another where the impurity is unscreened in the ground state but may be screened at higher energy scales. When $\alpha$ increases beyond $\frac{3}{2}\pi$, the impurity can not be screened at any energy scale. 

The bound mode energy $E_b$  being negative in the region $\alpha \le \pi$ and positive in the region $\alpha \ge \pi$, one might conclude on the basis of purely energetic considerations that the impurity is screened (resp. unscreened) in the region $\alpha \le \pi$ (resp. $\alpha \ge \pi$) with a first-order phase transition at $\alpha=\pi$ where the two states cross.   Such an argument resembles the YSR mechanism \cite{yu,Shiba,Rusinov,Kondosup} for the quantum phase transition between screened and local moment phases for a Kondo impurity coupled to a s-wave superconductor, even though here the system is gapless in the bulk \cite{pasnoori2022rise}.  However, dynamical considerations need to be applied in addition to energetics. Since $\Gamma_b < 0$    when $\alpha \le \pi$, the impurity is eventually found to be unscreened at sufficiently large time $1/\Gamma_0\gg t \gg \tau_b\equiv1/|\Gamma_b|$ even when $|B\rangle$ is lower in energy. Preparing the state of the system at time $t=0$ in a linear combination of the unscreened and screened states \footnote{Since the $\ket{B}$ and  $\ket{U}$ states do not have the same fermionic parity, one needs to add a spinon in bulk to (say) the $\ket{B}$ state.} $\ket{\psi} = u \ket{U}+ b \ket{B}$. 
After some time $t$ (assuming there have been no losses in the meantime due to the jump operators in the Lindbladian \cite{nakagawa2018non}) the wave function evolves as $\ket{\psi(t)}\sim u \ket{U} + b e^{-t/\tau_b}  \ket{B}$ with the system ending in the unscreened state $ \ket{U} $. Therefore, due to the appearance of the time scale $\tau_b$  in the $\widetilde{YSR}$, the phase transition between screened and unscreened phases takes place at $\alpha =\pi/2$. However, close enough to $\alpha =\pi/2$,
$\tau_b$ can be large  so we expect that the $\ket{B}$ state can  be observed before the whole system decays. As one departs far enough from $\alpha =\pi/2$, $\tau_b \sim T_K^{-1}$, 
a more accurate description of the system would require a better understanding of the intricate balance between energetics and dynamic stability in this problem.  This goes beyond the scope of the present work.

 Since it is experimentally viable to engineer a two-orbital system in a cold atom system\cite{riegger2018localized} with both localized and itinerant degrees of freedom, the experimental realization of the Kondo effect in out-of-equilibrium setting has become a possibility. We expect that by turning the optical frequency to modulate the rate of loss, one should be able to probe the transition from the Kondo phase to $\widetilde{YSR}$ phase and eventually to the local moment phase.  Moreover, by exploring the $\widetilde{YSR}$ phase, one may probe the intricate dynamics between the states in which impurity is screened and not screened.


\textit{Acknowledgements-}N.A. wishes to thank M. Schiro and H. Saleur for exciting discussions.

\bibliography{ref}
\onecolumngrid 

\newpage

\begin{appendix}

In these appendixes , we provide additional details regarding the solution of Bethe Ansatz equations in all three phases. Moreover, in the Kondo phase, we do a more careful analysis of Bethe Ansatz equations and derive an analytic equation for the imaginary part of the root distribution and also the imaginary part of the spinon energy.  
\section{Solution of the Bethe Ansatz Equations}
The Bethe Ansatz equations are

\begin{equation}
e^{i k_{j} L}=\prod_{\gamma=1}^{M} \frac{\Lambda_{\gamma}-1+i c / 2}{\Lambda_{\gamma}-1-i c / 2},
\end{equation}

and 

\begin{equation}
\prod_{\delta=1,\delta\neq \gamma}^{M} \frac{\Lambda_{\delta}-\Lambda_{\gamma}+i c }{\Lambda_{\delta}-\Lambda_{\gamma}-i c }=\left(\frac{\Lambda_{\gamma}-1-i c  / 2}{\Lambda_{\gamma}-1+i c  / 2}\right)^{N^{e}}\left(\frac{\Lambda_{\gamma}-1+e^{-i\phi}-i c  / 2}{\Lambda_{\gamma}-1+e^{-i\phi}+i c  / 2}\right).
\label{ssbae}
\end{equation}

Now, taking Log on both sides of the equation recalling $\ln \left(\frac{i+\frac{z}{c}}{i-\frac{z}{c}}\right)=i \Theta(z)$, we get
\begin{equation}
    i N^e\Theta(2(\Lambda_\gamma-1))+i\Theta(2(\Lambda_\gamma-1+e^{-i\phi})) =-2i\pi I_\gamma+i\sum_{\delta=1}^M\Theta(\Lambda_\gamma-\Lambda_\delta).
    \label{sbeth}
\end{equation}

The total energy can be written as
\begin{equation}
E=\sum_{j=1}^{N^e}k_j=\sum_{j=1}^{N^{e}} \frac{2 \pi}{L} n_{j}+D \sum_{\gamma=1}^{M}\left[\Theta\left(2 \Lambda_{\gamma}-2\right)-\pi\right],
\label{cbeth}
\end{equation}
where $D=\frac{N^e}{L}$ is the density of electrons.

Notice that $n_j$ is an integer, and $I_\gamma$ is an integer if $N-M-1$ is odd and a half-odd integer if $M$ is even. Each allowed choice of these numbers $\{n_j,I_\gamma\}$ uniquely determines an eigenstate of the Hamiltonian. Thus, we call these numbers the quantum numbers of the state they determine. It is remarkable that the momenta $k_j$, also called charge rapidities, do not appear in Eq.\eqref{sbeth}. Thus, these two equations Eq.\eqref{cbeth} and Eq.\eqref{sbeth} can be solved independently, which shows that the charge and the spin completely decouple.

We solve the Bethe equations in different parametric regimes.

\subsection{The Kondo Phase}
When $\sin(\phi)<\frac{c}{2}$, we can solve Eq.\eqref{sbeth} in the thermodynamic limit by computing the density of the roots defined as $\sigma(\Lambda)=\frac{1}{\Lambda_{\gamma+1}-\Lambda_\gamma}$ describing the number of solution in the interval $(\Lambda,\Lambda+d\Lambda)$ of the solution rather than the solution $\Lambda_\gamma$ themselves. In terms of the density, Eq.\eqref{sbeth} can be written as
\begin{equation}
N^{e} \Theta\left(2 \Lambda_{\gamma}-2\right)+\Theta\left(2 (\Lambda_{\gamma}-1+e^{-i\phi}\right)=\int_C d \Lambda^{\prime} \sigma\left(\Lambda^{\prime}\right) \Theta\left(\Lambda_{\gamma}-\Lambda^{\prime}\right)-2 \pi I_{\gamma},
\label{cpmom}
\end{equation}
where the integration is over the locus of the Bethe equation, which lies in the complex plane.
By subtracting the above equation for solution $\Lambda_\gamma$ and $\Lambda_{\gamma+1}$ and expanding in the difference $\Delta\Lambda=\Lambda_{\gamma+1}-\Lambda_\gamma$, the ground state density $\sigma_0(\Lambda)$ can be written 
\begin{equation}
\sigma_0(\Lambda)=f(\Lambda)-\int_C K(\Lambda-\Lambda')\sigma_0(\Lambda')\mathrm d\Lambda',
\label{densdens}
\end{equation}
where
\begin{equation}
f(\Lambda)=\frac{2c}{\pi}\left[\frac{N^e}{c^2+4(\Lambda-1)^2}+\frac{1}{c^2+4(\Lambda-1+e^{-i\phi})^2} \right] \quad\text{and}\quad
K(\Lambda)=\frac{1}{\pi}\frac{c}{c^2+\Lambda^2}.
\end{equation}

As we show below, the density of the roots is different depending on the relation between c and $\phi$. 

In the Hermitian case, the integral over $C$ is the integral over the real line, but here, the locus of the Bethe equation deviates from the real line. However, in terms of the $\Lambda$ variables, the deviation is of order $\frac{1}{N}$ and since the integrand $K(\Lambda-\Lambda')\sigma_0(\Lambda)$ is analytic, we can deform this integration to the real line.

The above equation can be solved in the Fourier space 
\begin{equation}
    \tilde\sigma_0(p)=\frac{1}{2} e^{-i p } N^e \text{sech}\left(\frac{c p }{2}\right)+\frac{1}{2}  e^{i p  e^{-i \phi }-i p } \text{sech}\left(\frac{c p }{2}\right),
\end{equation}
which can be written in the $\Lambda-$space via inverse Fourier transform as
\begin{equation}
\sigma_0(\Lambda)=\frac{1}{2 c}\left[N^e \operatorname{sech}\left(\frac{\pi  (\Lambda-1)}{c}\right)+\operatorname{sech}\left(\frac{\pi ( \Lambda-1+e^{-i\phi}}{c}\right)\right].
\end{equation}

Thus, the ground state magnetization can be computed as
\begin{equation}
S=\frac{N^e+1}{2}-\int\sigma_0(\Lambda)\mathrm{d}\Lambda=\frac{N^e+1}{2}-\left[\frac{N^e}{2}+\frac{1}{2} \right]=0.
\end{equation}

The energy of the ground state can be computed as
\begin{align}
    E_0&=-\frac{\pi N^e\left(N^e+1\right)}{L}-\frac{N^e}{L} \pi {\int_{-\infty}^{\infty} \mathrm{d} \Lambda \sigma_0(\Lambda)}+D \int \mathrm{d} \Lambda \sigma_0(\Lambda) \Theta(2 \Lambda-2)\nonumber\\
    &=-\frac{\pi}{2 L}\left(N^e\right)^2-i D \log \left(\frac{\Gamma \left(\frac{1}{2}-\frac{i e^{-i \phi }}{2 c}\right) \Gamma \left(1+\frac{i e^{-i \phi }}{2 c}\right)}{\Gamma \left(1-\frac{i e^{-i \phi }}{2 c}\right) \Gamma \left(\frac{1}{2}+\frac{i e^{-i \phi }}{2 c}\right)}\right) .
\end{align}

\paragraph{Elementary Excitations}
The variation from the ground state configuration $\{n_j^0,I_\gamma^0\}$ gives the elementary excitations of the model. There are two types of excitations
\begin{enumerate}[label=(\alph*)]
\item \underline{Charge excitations:} There are the excitations obtained by exciting the charge degrees of freedom. When we change a given $n_{j}$ where $-K \leq(2 \pi / L) n_{j}<0$ to $n_{j}^{\prime}=n_{j}+\Delta n \geq 0 .$, energy is changed by
\begin{equation}
\Delta E=\frac{2 \pi}{L} \Delta n>0.
\end{equation}
This once again shows that the charge degree of the freedom is completely decoupled from the spin degree as the above change does not change $M$, which depends only on quantum numbers $\{I_\gamma\}$.
\item \underline{Spin excitations}: There are obtained by altering the sequence $\{I_\gamma^0\}$ from the ground state configuration without changing the quantum numbers $n_j$. We can vary the configuration by putting ``holes" into it, where a ``hole" means an integer omitted from the consecutive sequence. In the presence of the hole, the  density given by Eq.\eqref{densdens} becomes
\begin{equation}
\sigma(\Lambda)+\sigma^{h}(\Lambda)=f(\Lambda)-\int K\left(\Lambda-\Lambda^{\prime}\right) \sigma\left(\Lambda^{\prime}\right) d \Lambda^{\prime},
\end{equation}
where the hole density is given by
\begin{equation}
\sigma^{h}(\Lambda)=\sum_{i=1}^{N^{h}} \delta\left(\Lambda-\Lambda_{i}^{h}\right).
\end{equation}
Once again, we use the Fourier transformation to write the solution. This time, we write the solution in the Fourier space
\begin{equation}
\Delta\tilde\sigma(p)=\tilde{\sigma}(p)-\tilde{\sigma}_{0}(p)=-\sum_{j}^{N^h} e^{-i\Lambda_j^h p} \frac{1}{e^{-c | p| }+1}=-\sum_{j}^{N^h}\frac{ e^{-i\Lambda_j^h p+ \frac{c}{2}|p|}}{2\cosh \left( \frac{cp}{2} \right)}.
\end{equation}
Because all the momenta $\Lambda$ are coupled through Eq.\eqref{cpmom}, removing one of them affects all as suggested by the dressing of hole density in Fourier space $e^{-i\Lambda_j^h p}$ to $\frac{ e^{-i\Lambda_j^h p+ \frac{c}{2}|p|}}{2\cosh \left( \frac{cp}{2} \right)}$.
The total number of down spins can be computed by taking the integral of the density
\begin{equation}
M=\int \sigma(\Lambda)\mathrm{d}\Lambda=\left.\tilde{\sigma}(p)\right|_{p=0}=\frac{N}{2}-\frac{N^h}{2}.
\end{equation}
Thus, the contribution due to a single hole is $(\Delta M)_h=-\frac{1}{2}$. 
\end{enumerate}

\paragraph{The triplet excitation}{\phantom{.}}\\
Now, we consider the simplest excitation made up of two holes at say $\Lambda_1^h$ and $\Lambda_2^h$. The excitation energy is given by
\begin{equation}
\Delta E^t=D\int \Delta\sigma(\Lambda)\left[\Theta(2\Lambda-2)-\pi\right]\mathrm{d}\Lambda=2 D\left(\tan ^{-1} e^{(\pi / c)\left(\Lambda_{1}^{h}-1\right)}+\tan ^{-1} e^{(\pi / c)\left(\Lambda_{2}^{h}-1\right)}\right),
\end{equation}
which shows that this is a sum of two terms carrying spin-$\frac{1}{2}$, which gives a total spin-one state. \\

To get a spin one-half state, we need to add a hole with an electron, which gives a total energy
\begin{equation}
\Delta E^d=2 D\left(\tan ^{-1} e^{(\pi / c)\left(\Lambda_{1}^{h}-1\right)}\right)+\frac{2\pi}{L}n,
\end{equation}
where the first term is the energy of the excitation carrying only the spin degree of freedom (spinons), and the second term is the energy of excitation carrying only the charge degree of freedom (holons).

\paragraph{The singlet excitation}{\phantom{.}}\\
So far, we have only looked at the continuous roots of the Bethe equation and studied the perturbation around it by changing the quantum numbers $n_j$ and $I_\gamma$. In order to obtain the singlet excitation, we need to look at the discrete complex roots of the Bethe equation. 

Adding two holes and a string solution, we write
\begin{equation}
\sigma(\Lambda)+\sigma^h(\Lambda)=f(\Lambda)-\int K\left(\Lambda-\Lambda^{\prime}\right) \sigma\left(\Lambda^{\prime}\right) d \Lambda^{\prime}-\sigma^{s t}(\Lambda),
\end{equation}
where

\begin{align}
\sigma^h(\Lambda) & =\delta\left(\Lambda-\Lambda_1^h\right)+\delta\left(\Lambda-\Lambda_2^h\right) \\
\sigma^{s t}(\Lambda) & =K_3(\Lambda-\bar{\Lambda})+K_1(\Lambda-\bar{\Lambda}) .
\end{align}

The second equation which fixes the position of the 2-string $\bar{\Lambda}$ is of the form

\begin{equation}
 N^e \Theta(\bar{\Lambda}-1)+\Theta(\bar{\Lambda}-1+e^{-i\phi})=-2 \pi I^{(2)}+\sum_{\delta=1}^M \Theta_1\left(\bar{\Lambda}-\Lambda_\delta\right)+\sum_{\delta=1}^M \Theta_3\left(\bar{\Lambda}-\Lambda_\delta\right).
 \label{stposeqn}
\end{equation}

It can be shown that $\bar\Lambda=\frac{1}{2}(\Lambda_1^h+\Lambda_2^h)$ and in the thermodynamic limit, the energy of the singlet is equal to the energy of the triplet excitations.

Thus, starting from the ground state, all the excited states are constructed by exciting the charge degree of freedom, adding an even number of spinons, or adding string solutions with an appropriate number of spinons.

\paragraph{Analytical expression for the imaginary part of roots positions and energies}
By using the expressions for the density of roots, one can decouple the BAE \eqref{ssbae} by writing the LHS as 
\begin{equation}
    \begin{aligned}
        \prod_{\delta=1}^{M} \frac{\Lambda_{\delta}-\Lambda_{\gamma}+i c }{\Lambda_{\delta}-\Lambda_{\gamma}-i c } =& \exp\left(\int \mathrm{d} \Lambda' \sigma(\Lambda') \log\frac{\Lambda'-\Lambda_{\gamma}+i c }{\Lambda'-\Lambda_{\gamma}-i c }\right)\\
        =&\exp\left[\int_0^\infty \frac{\mathrm{d} p}{p} e^{-cp}\left(\Tilde{\sigma}(p)e^{i\Lambda_{\gamma} p} - \Tilde{\sigma}(-p)e^{-i\Lambda_{\gamma} p}\right)+i\pi \Tilde{\sigma}(0)\right]\\
        =&\exp[N^e Q^{\mathrm{bulk}}+Q^{\mathrm{imp}}+ \sum_{k=1}^h Q^h_k].
    \end{aligned}
\end{equation}
To get $Q$s we will be using the following integral relation 
\begin{equation}
    \int_{\epsilon}^\infty \frac{\mathrm{d} p}{p} \frac{e^{-ap}}{1+e^{-cp}}e^{ip\Lambda} = \kappa(\epsilon,c) + \ln \frac{\Gamma(\frac{a-i\Lambda}{2c})}{\Gamma(\frac{a+c-i\Lambda}{2c})},
\end{equation}
where $\kappa(\epsilon)$ is a term depending only on the cutoff and $c$ that logarithmically diverges as $\epsilon\rightarrow 0$. This yields
\begin{equation}
    \begin{aligned}
        Q^{\mathrm{bulk}}&=\log \frac{\Lambda_{\gamma}-1-ic/2}{\Lambda_{\gamma}-1+ic/2}\frac{e^{-\pi(\Lambda_{\gamma}-1)/c}-i}{e^{-\pi(\Lambda_{\gamma}-1)/c}+i}\\
        Q^{\mathrm{imp}}&=\log \frac{\Lambda_{\gamma}-1+e^{-i\phi}-ic/2}{\Lambda_{\gamma}-1+e^{-i\phi}+ic/2}\frac{e^{-\pi(\Lambda_{\gamma}-1)/c}-ie^{\pi e^{-i\phi}/c}}{e^{-\pi(\Lambda_{\gamma}-1)/c}+ie^{\pi e^{-i\phi}/c}}\\
        Q_j^h&=\log i\frac{\Gamma(\frac{c+i(\Lambda_{\gamma}-\Lambda^h_j)}{2c})}{\Gamma(\frac{c-i(\Lambda_{\gamma}-\Lambda^h_j)}{2c})}\frac{\Gamma(\frac{2c-i(\Lambda_{\gamma}-\Lambda^h_j)}{2c})}{\Gamma(\frac{2c+i(\Lambda_{\gamma}-\Lambda^h_j)}{2c})}.
    \end{aligned}
\end{equation}
Collecting terms, we arrive at the decoupled BAE:
\begin{equation}
    \begin{aligned}
        1=&-\left( \frac{e^{-\pi(\Lambda_{\gamma}-1)/c}-i}{e^{-\pi(\Lambda_{\gamma}-1)/c}+i}\right)^{N^e}\frac{e^{-\pi(\Lambda_{\gamma}-1)/c}-ie^{\pi e^{-i\phi}/c}}{e^{-\pi(\Lambda_{\gamma}-1)/c}+ie^{\pi e^{-i\phi}/c}} \prod_j^h \frac{\Gamma(\frac{c+i(\Lambda_{\gamma}-\Lambda^h_j)}{2c})}{\Gamma(\frac{c-i(\Lambda_{\gamma}-\Lambda^h_j)}{2c})}\frac{\Gamma(\frac{2c-i(\Lambda_{\gamma}-\Lambda^h_j)}{2c})}{\Gamma(\frac{2c+i(\Lambda_{\gamma}-\Lambda^h_j)}{2c})},
    \end{aligned}
\end{equation}
or simpler
\begin{equation}
    \begin{aligned}
        1=&e^{-i2N^e \tan^{-1}(e^{\pi(\Lambda_{\gamma}-1)/c})}e^{iq(\Lambda_{\gamma},\phi)-p(\Lambda_{\gamma},\phi)} \prod_j^h S^{-1}(\Lambda_{\gamma}-\Lambda^h_j),
    \end{aligned}
\end{equation}
 where we obtained a well-known expression of the physical S-matrix of spinons 
\begin{equation}
    S(\Lambda^h)=i\frac{\Gamma(\frac{1}{2}-i\Lambda^h/2c)}{\Gamma(\frac{1}{2}+i\Lambda^h/2c)}\frac{\Gamma(1+i\Lambda^h/2c)}{\Gamma(1-i\Lambda^h/2c)},
\end{equation}
and simplified the impurity term to
\begin{equation}
    e^{iq(\Lambda_{\gamma},\phi)-p(\Lambda_{\gamma},\phi)}\equiv\frac{e^{-\pi(\Lambda_{\gamma}-1)/c}-ie^{\pi e^{-i\phi}/c}}{e^{-\pi(\Lambda_{\gamma}-1)/c}+ie^{\pi e^{-i\phi}/c}}.
\end{equation}
Let's consider the ground state with no holes and only 1-string roots. Then the one-string roots $\Lambda_{\gamma}$ should satisfy the equation
\begin{equation}\begin{aligned}
        1=&e^{-i2N^e \tan^{-1}(e^{\pi(\Lambda_{\gamma}-1)/c})}e^{iq(\Lambda_{\gamma},\phi)-p(\Lambda_{\gamma},\phi)}.
    \end{aligned}
\end{equation}
Writing the 1-string explicitly as $\Lambda_{\gamma}=\mu+i\nu$ (where we make an assumption that later proves to be self-consistent that the imaginary part is of the order $O(1/N^e)$), this equation simplifies to real and imaginary parts
\begin{equation}
   N^e\left[ 2\tan^{-1}(e^{\pi(\mu-1)/c})-\frac{1}{N^e}q(\mu,\phi)\right] =2\pi J(\mu)
\end{equation}
\begin{equation}
    \frac{\pi N^e}{c \cosh(\pi (\mu-1)/c)}\nu(\mu)=p(\mu,\phi).
\end{equation}
The first equation agrees with the one we started with at the beginning of this section. The latter gives the imaginary part of the positions of 1-strings:
\begin{equation}
\nu(\mu)=\frac{c \cosh(\pi(\mu-1)/c)}{ 2\pi N^e}\log \frac{\cosh(\frac{\pi(\mu-1+\cos\phi)}{c})+\sin(\frac{\pi\sin\phi}{c})}{\cosh(\frac{\pi(\mu-1+\cos\phi)}{c})-\sin(\frac{\pi\sin\phi}{c})}.
    \label{comploc}
\end{equation}

\paragraph{Imaginary part of the spinon energy}Taking into account that the holes lie along the 1-string curve taking complex values $\Lambda^h=\mu_h + \frac{i}{N^e}\nu(\mu_h)$ we get the spinon energy $\mathcal{E}_h=2 D\tan ^{-1} e^{(\pi / c)\left(\Lambda^h-1\right)}$ to be consisting of the real $E_h$ and imaginary $\Gamma_h$ parts $\mathcal{E}_h=E_h +i \Gamma_h$ with
\begin{equation}
    \begin{aligned}
        E_h &=2 D\tan ^{-1} e^{(\pi / c)\left(\mu_h-1\right)}\\
        \Gamma_h &= \frac{\pi N^e}{c L \cosh(\pi(\mu_h-1)/c)}  \nu(\mu_h) = \frac{1}{2L}\log \frac{\cosh(\frac{\pi(\mu_h-1+\cos\phi)}{c})+\sin(\frac{\pi\sin\phi}{c})}{\cosh(\frac{\pi(\mu_h-1+\cos\phi)}{c})-\sin(\frac{\pi\sin\phi}{c})}
    \end{aligned}
\end{equation}
the latter is always positive. Now let's consider how this expression behaves in the scaling limit \(c \to 0\) while keeping \(\alpha = \dfrac{\pi \sin \phi}{c}\) fixed and using universal $\theta=\frac{\pi}{c}\Lambda_h$. This results in the decay width being
\begin{equation}
\Gamma_h(\theta) = \frac{1}{2L}\log \frac{\cosh \theta+\sin \alpha}{\cosh \theta-\sin \alpha}.
\end{equation}
These results show that the imaginary part of the spinon energy scales as \(1/L\). Alternatively, one can write this in terms of the real part of the spinon energy $E_h$
\begin{equation}
    \Gamma_h(E_h)=\frac{1}{L}\;  \tanh^{-1} \left(\frac{2E_hT_K}{E_h^2+T_K^2} \sin \alpha\right)
\end{equation}
which defines a complex curve $\mathcal{E}(E)=E+i\Gamma(E)$ formed by the spinon energies.

\paragraph{Density of states}
The spinon energies form the complex curve $\mathcal{E}(E)=E+i\Gamma(E)$, and now one can define a complex-valued density of states (DOS) along that curve as $\rho(\mathcal{E})=\frac{\mathrm{d} N}{\mathrm{d} \mathcal{E}}$ 
\begin{equation}
    \rho(\mathcal{E})=\sigma_0(\Lambda)/\frac{\mathrm{d}\mathcal E}{\mathrm{d}\Lambda}=\frac{L}{2\pi}+\frac{ T_0}{\pi  \left(\mathcal E ^2+ T_0^2\right)},
\end{equation}
where $T_0=T_K e^{i\alpha}$. Let's define a real-valued DOS through the number of states $\mathrm{N}$ lying within the range of real energies $\mathrm{d} E$:  $\tilde{\rho}(E) = \frac{\mathrm{d} N}{\mathrm{d} E}$. Then we have
$$
\rho(\mathcal{E}) d \mathcal{E}=\rho(E+i \Gamma(E))\left(1+i \partial_E \Gamma(E) \right) d E \equiv \tilde{\rho}(E) d E .
$$
This yields 
$$
 \tilde{\rho}(E) = \frac{L}{2\pi}+\Re\left(\frac{ T_0}{\pi  \left(\mathcal E ^2+ T_0^2\right)}\right)+i\left[ \Im\left(\frac{ T_0}{\pi  \left(\mathcal E ^2+ T_0^2\right)}\right)+\frac{L}{2\pi} \partial_E \Gamma(E) \right]+O\left(\frac{1}{L}\right)
$$
with the imaginary part canceling exactly. The final expression for the real-valued DOS is
\begin{equation}
    \tilde{\rho}(E) =\frac{L}{2\pi}+\Re\left(\frac{ T_0}{\pi  \left(\mathcal E ^2+ T_0^2\right)}\right) +O\left(\frac{1}{L}\right) =\Re\left(\rho(E)\right) +O\left(\frac{1}{L}\right)\;.
\end{equation}

$$
$$



\subsection{Bound mode phase}
When $\frac{c}{2}<\sin(\phi)<\frac{3c}{2}$, there is a new solution of the Bethe equation in the thermodynamic limit of the form
\begin{equation}
    \Lambda_{IS}=1-\cos(\phi)+ \frac{i}{2}(2\sin(\phi)-c).
    \label{addsln1}
\end{equation}

In this regime, the solution of Eq.\eqref{densdens}, which gives the distribution of the continuous root distribution, can be written in the Fourier space as
\begin{equation}
\tilde\sigma_0(p)=\frac{1}{2} e^{-i p } N^e \text{sech}\left(\frac{c p }{2}\right)+\frac{\left(e^{c p}-1\right) \theta (-p) e^{-\frac{1}{2} p (c-2 \sin (\phi )-2 i \cos (\phi )+2 i)}}{e^{-c | p| }+1}.
\label{nkgsomsp}
\end{equation}

We compute 
\begin{equation}
M=\int_{-\infty}^\infty \sigma_0(\Lambda)\mathrm{d}\Lambda=\tilde\sigma_0(0)=\frac{N^e}{2}=\frac{N-1}{2}.
\end{equation}
Since for $N$ even or $N^e$ odd, the number of $M$ is not an integer. Thus, we need to add a hole or the impurity string solution. Adding a hole, we get
\begin{equation}
    \sigma_0(\Lambda)+\delta(\Lambda-\Lambda_h)=f(\Lambda)-\int K(\Lambda-\Lambda')\sigma_0(\Lambda')\mathrm{d}\Lambda' ,
\end{equation}

where
\begin{equation}
    K(\Lambda)=\frac{1}{\pi}\frac{c}{c^2+\Lambda^2},
\end{equation}
and
\begin{equation}
   f(\Lambda)=\frac{2c}{\pi}\left[\frac{N^e}{c^2+4(\Lambda-1)^2}+\frac{1}{c^2+4(\Lambda-1+e^{-i\phi})^2} \right] .
\end{equation}

 Now, we solve the above equation in Fourier space such that
\begin{equation}
\tilde\sigma_1(p)=\frac{1}{2} e^{-i p } N^e \text{sech}\left(\frac{c p }{2}\right)+\frac{\left(e^{c p}-1\right) \theta (-p) e^{-\frac{1}{2} p (c-2 \sin (\phi )-2 i \cos (\phi )+2 i)}}{e^{-c | p| }+1}-\frac{e^{-i p \Lambda _h}}{e^{-c | p| }+1},
\end{equation}
such that
\begin{equation}
    M=\tilde\sigma_0(p)=\frac{1}{2} \left(N^e-1\right)=\frac{N-2}{2}.
   \label{gstmval}
\end{equation}
Hence, the spin in this state is
\begin{equation}
    S=\frac{N}{2}-M=1.
\end{equation}
This is the state where impurity is unscreened. The unscreened spin and the hole make a triplet pairing. 

The energy of the state described by the continuous root distribution is
\begin{align}
  E_{ar}&=  -\frac{\pi}{2 L}\left(N^e\right)^2+D\int \mathrm{d}p \frac{\left(e^{c p}-1\right) \theta (-p) e^{-\frac{1}{2} p (c-2 \sin (\phi )-2 i \cos (\phi )+2 i)}}{e^{-c | p| }+1}\left(-\frac{i e^{-\frac{c | p| }{2}+i p}}{p} \right)\nonumber\\
  &=-\frac{\pi}{2L}(N^e)^2-i D \log \left(\frac{\Gamma \left(\frac{i e^{-i \phi }}{2 c}\right) \Gamma \left(1+\frac{i e^{-i \phi }}{2 c}\right)}{\Gamma \left(\frac{c+e^{-i \phi } i}{2 c}\right)^2}\right).
  \label{engareqn}
\end{align}

Adding string solution Eq.\eqref{addsln1}, we write
\begin{align}
    &\left(\frac{\Lambda_{\gamma}-1+e^{-i\phi}+i c  / 2}{\Lambda_{\gamma}-1+e^{i\phi}-i c  / 2}\right) \left(\frac{\Lambda _{\gamma }+\cos (\phi )-1-i \left(\frac{c}{2}+\sin (\phi )\right)}{\Lambda _{\gamma }+\cos (\phi )-1+i \left(\frac{3 c}{2}-\sin (\phi )\right)}\right)  \prod_{\delta=1,\delta\neq \gamma}^{M} \frac{\Lambda_{\delta}-\Lambda_{\gamma}+i c }{\Lambda_{\delta}-\Lambda_{\gamma}-i c }=\left(\frac{\Lambda_{\gamma}-1-i c  / 2}{\Lambda_{\gamma}-1+i c  / 2}\right)^{N^{e}}.
    \label{fundstring}
\end{align}
Taking Log on both sides of the equation, we obtain

\begin{align}
   N^{e} \Theta\left(2 \Lambda_{\gamma}-2\right)&+\Theta\left(2 (\Lambda_{\gamma}-1+e^{-i\phi}\right)=i \log \left(\frac{\Lambda _{\gamma }+\cos (\phi )-1-i \sin (\phi )-\frac{i c}{2}}{\Lambda _{\gamma }+\cos (\phi )-1-i \sin (\phi )+\frac{3 i c}{2}}\right)+\int d \Lambda^{\prime} \sigma\left(\Lambda^{\prime}\right) \Theta\left(\Lambda_{\gamma}-\Lambda^{\prime}\right)-2 \pi J_{\gamma}.
    \label{st1}
\end{align}
Changing the sum to the integral as usual, we obtain
\begin{multline}
    -\frac{4 c N^e}{c^2+4 \left(\Lambda -1\right){}^2}-\frac{4 c}{c^2+4 \left(\Lambda +e^{-i \phi }-1\right){}^2}+\frac{i}{\frac{3 i c}{2}+\Lambda +e^{-i \phi }-1}-\frac{i}{-\frac{i c}{2}+\Lambda +e^{-i \phi }-1}\\=-2\pi\sigma(\Lambda)-\int \sigma(\Lambda')\frac{2c}{c^2+(\Lambda-\Lambda')^2}\mathrm{d}\Lambda' .
\end{multline}

Solving the above equation in Fourier space, we obtain the contribution from the impurity string solution as
\begin{equation}
    \Delta\tilde\sigma^{\mathrm{ist}}(p)=\begin{cases}
        -\frac{e^{-\frac{c p}{2}+i p e^{-i \phi }-i p}}{e^{c | p| }+1}\quad\quad\quad\quad\quad\quad\quad \text{when } 3 c>2 \sin (\phi )\\
       -\frac{\left(e^{2 c p}-1\right) \theta (-p) e^{-\frac{3 c p}{2}+i p e^{-i \phi }-i p}}{e^{-c | p| }+1}  \quad \text{when } 3 c<2 \sin (\phi ). \\
    \end{cases}
\end{equation}

When $3c>2\sin(\phi)$, we compute
\begin{equation}
 \Delta M=  1+\int \Delta \sigma^{\mathrm{ist}}(\Lambda)\mathrm{d}\Lambda= 1+\Delta\tilde\sigma^{\mathrm{ist}}(0)=\frac12.
\end{equation}
\\

The energy of the string solution when $3c>2\sin(\phi)$  is calculated as
\begin{align}
    \Delta E^{\mathrm{ist}}&={D}\int \Delta\sigma^{\mathrm{ist}}(\Lambda)[\Theta(2\Lambda-2)-\pi]\mathrm{d}\Lambda+D\left(\Theta\left(2 \Lambda_{stp}^{+}-2\right)- \pi\right)\nonumber\\
      &=D\frac{\pi }{2}+i D \log \left(-\frac{i c \left(1+e^{\frac{\pi  e^{-i \phi }}{c}}\right) \left(2-\frac{2 i e^{-i \phi }}{c}\right)}{\left(-1+e^{\frac{\pi  e^{-i \phi }}{c}}\right) \left(-2 c+2 i e^{-i \phi }\right)}\right)\nonumber\\
      &=iD \log \left(\coth \left(\frac{\pi  e^{-i \phi }}{2 c}\right)\right).
      \label{bndryengeqn}
\end{align}
Upon taking the scaling limit, the energy can be written as
\begin{equation}
  \Delta E^{\mathrm{ist}}=  iT_K  e^{i\pi \alpha}.
\end{equation}
Thus, there are two unique kinds of states in this phase. Adding a hole on top of the continuous root distribution, we obtain the state where the impurity is unscreened. This state has energy
\begin{equation}
    E_{u}=-\frac{\pi}{2L}(N^e)^2-i D \log \left(\frac{\Gamma \left(\frac{i e^{-i \phi }}{2 c}\right) \Gamma \left(1+\frac{i e^{-i \phi }}{2 c}\right)}{\Gamma \left(\frac{c+e^{-i \phi } i}{2 c}\right)^2}\right)+E_\theta ,
\end{equation}
where $E_\theta=2 D\left(\tan ^{-1} e^{(\pi / c)\left(\Lambda_{1}^{h}-1\right)}\right)$ is the spinon energy.

The other state is obtained by adding the impurity string solution on top of the continuous root distribution. In this state, the impurity is screened by the bound mode formed at the impurity site. The energy of this state is
\begin{equation}
    E_b=-\frac{\pi}{2L}(N^e)^2-i D \log \left(\frac{\Gamma \left(\frac{i e^{-i \phi }}{2 c}\right) \Gamma \left(1+\frac{i e^{-i \phi }}{2 c}\right)}{\Gamma \left(\frac{c+e^{-i \phi } i}{2 c}\right)^2}\right)+iD \log \left(\coth \left(\frac{\pi  e^{-i \phi }}{2 c}\right)\right).
\end{equation}

All the excited states are constructed by adding charge excitations, even the number of spinons, string solutions, etc., on top of these two states. 

Notice that the impurity string solution has a real part of the energy that is negative when $c/2\sin\phi<c$ and positive when $c<\sin\phi<3c/2$.

Since the root distribution is the same in the Kondo and bound mode phase when the impurity string solution is added, the impurity density of state in the state with the bound mode is an analytic continuation of the DOS in the Kondo phase \textit{i.e.}
\begin{equation}
    \rho(\mathcal{E})=\frac{T_0}{\pi  \left(\mathcal{E} ^2+T_0^2\right)},
\end{equation}
and the imaginary part of the spinon energy is also given by the same expression as in the Kondo phase \textit{i.e.}
\begin{equation}
    \Gamma(E)=\tanh ^{-1}\left(\frac{2 E \sin (\alpha ) T_K}{E^2+T_K^2}\right).
\end{equation}

In the state where the impurity string is not added, the impurity contribution to the root distribution is
\begin{equation}
  \sigma_{\mathrm{imp}}(\Lambda)=  -\frac{1}{\pi  \left(c-2 i \Lambda -2 i e^{-i \phi }+2 i\right)}-\frac{\psi ^{(0)}\left(\frac{1}{4}-\frac{i}{2 c}+\frac{i e^{-i \phi }}{2 c}+\frac{i \Lambda }{2 c}\right)}{2 \pi  c}+\frac{\psi ^{(0)}\left(-\frac{1}{4}-\frac{i}{2 c}+\frac{i e^{-i \phi }}{2 c}+\frac{i \Lambda }{2 c}\right)}{2 \pi  c}.
\end{equation}

Hence, the Density of states is
\begin{equation}
  \rho(\mathcal{E})=\frac{-\frac{2 \pi }{\pi -2 i \log \left(\frac{E}{T_0}\right)}-\psi ^{(0)}\left(\frac{2 i \log \left(\frac{E}{T_0}\right)+\pi }{4 \pi }\right)+\psi ^{(0)}\left(\frac{i \log \left(\frac{E}{T_0}\right)}{2 \pi }-\frac{1}{4}\right)}{2 \pi ^2 E}.
\end{equation}

and the imaginary part of the spinon energy is given by

\begin{equation}
    \Gamma(E)=\frac{1}{2} \log \left(16 \alpha ^2-8\pi  \alpha +4 \log ^2\left(\frac{E}{T_K}\right)+\pi ^2\right)+\log \left(\frac{\Gamma \left(\frac{\alpha }{2 \pi }-\frac{1}{4}-\frac{i \log \left(\frac{E}{T_K}\right)}{2 \pi }\right) \Gamma \left(\frac{\alpha }{2 \pi }+\frac{i \log \left(\frac{E}{T_K}\right)}{2 \pi }-\frac{1}{4}\right)}{\Gamma \left(\frac{\alpha }{2 \pi }+\frac{1}{4}-\frac{i \log \left(\frac{E}{T_K}\right)}{2 \pi }\right) \Gamma \left(\frac{\alpha }{2 \pi }+\frac{i \log \left(\frac{E}{T_K}\right)}{2 \pi }+\frac{1}{4}\right)}\right)-\log(4\pi).
\end{equation}

Adding the impurity string solution, the Bethe Ansatz equation can be written as
\begin{equation}\left(\frac{\Lambda_{\gamma}-1+\cos\phi-i\sin\phi+i \frac{c}{2}}{\Lambda_\gamma-1+\cos(\phi)-i\sin(\phi)+i\frac{3}{2}c}\right) \prod_{\delta=1,\delta\neq \gamma}^{M} \frac{\Lambda_{\delta}-\Lambda_{\gamma}+i c}{\Lambda_{\delta}-\Lambda_{\gamma}-i c  }=\left(\frac{\Lambda_{\gamma}-1-i c  / 2}{\Lambda_{\gamma}-1+i c  / 2}\right)^{N^{e}} ,
\end{equation}

which gives us a new solution
\begin{equation}
   HIS=1-\cos(\phi)+i\sin(\phi)-i\frac{3}{2}c,
\end{equation}
which we call higher-order impurity string solution (HIS).

 Upon adding the higher order boundary string and taking log on both sides, the Bethe equation becomes
\begin{equation}
N^e \Theta\left(2 \Lambda_\gamma-2\right)+\Theta\left(2\left(\Lambda_\gamma-1+e^{-i \phi}\right)\right)=i \log \left(\frac{\Lambda_\gamma+\cos (\phi)-1-i \sin (\phi)+\frac{i c}{2}}{\Lambda_\gamma+\cos (\phi)-1-i \sin (\phi)+\frac{5 i c}{2}}\right)+\int d \Lambda^{\prime} \sigma\left(\Lambda^{\prime}\right) \Theta\left(\Lambda_\gamma-\Lambda^{\prime}\right)-2 \pi J_\gamma,
\end{equation}
which gives an integral equation for the solution density as
\begin{equation}
-\frac{4 c N^e}{c^2+4(\Lambda-1)^2}-\frac{4 c}{c^2+4\left(\Lambda+e^{-i \phi}-1\right)^2}+\frac{i}{\frac{5 i c}{2}+\Lambda+e^{-i \phi}-1}-\frac{i}{\frac{i c}{2}+\Lambda+e^{-i \phi}-1}=-2 \pi \sigma(\Lambda)-\int \sigma\left(\Lambda^{\prime}\right) \frac{2 c}{c^2+\left(\Lambda-\Lambda^{\prime}\right)^2} \mathrm{~d} \Lambda^{\prime},
\end{equation}

such that the change in the solution density due to the higher-order string solution is
\begin{equation}
\Delta\tilde\sigma^{his}(p)=-\frac{e^{-\frac{3 c p}{2}+p \sin (\phi )+i p \cos (\phi )-i p}}{e^{c | p| }+1}.
\end{equation}

Now, the energy of this solution is
\begin{align}
\Delta E^{\mathrm{HIS}} & =D \int \Delta \sigma^{\mathrm{his}}(\Lambda)[\Theta(2 \Lambda-2)-\pi] \mathrm{d} \Lambda+D\left(\Theta\left(2 \Lambda_{his}-2\right)-\pi\right)\nonumber \\
& =-D \pi \Delta \sigma^{\mathrm{his}}(p=0)+D \int \mathrm{d} p\left(-\frac{e^{-\frac{3 c p}{2}+i p e^{-i \phi }-i p}}{e^{c | p| }+1}\right)\left(-\frac{i e^{-\frac{c|p|}{2}+i p}}{p}\right)+2 i D \tanh ^{-1}\left(3-\frac{2 i e^{-i \phi}}{c}\right)-\pi D\nonumber\\
&=D\int\mathrm{d}p \left(-\frac{e^{-\frac{3 c p}{2}+i p e^{-i \phi }-i p}}{e^{c | p| }+1}\right)\left(-\frac{i e^{-\frac{c|p|}{2}+i p}}{p}\right)-\frac{1}{2} (\pi  D)+2 i D \tanh ^{-1}\left(3-\frac{2 i e^{-i \phi}}{c}\right)\nonumber\\
&=i \log \left(\tanh \left(\frac{\pi  e^{-i \phi }}{2 c}\right)\right).
\end{align}

This is a remarkable result that the energy of the higher impurity sting solution is exactly negative of the energy of the fundamental boundary string. 

If we add the impurity string, the higher order impurity string, and then a hole, the solution density would then be of the form
\begin{align}
  \tilde \sigma(p)&  =\frac{1}{2} e^{-i p} N^e \operatorname{sech}\left(\frac{c p}{2}\right)+\frac{\left(e^{c p}-1\right) \theta(-p) e^{-\frac{1}{2} p(c-2 \sin (\phi)-2 i \cos (\phi)+2 i)}}{e^{-c|p|}+1}\nonumber\\
  &-\frac{e^{-\frac{c p}{2}+i p e^{-i \phi}}-i p}{e^{c|p|}+1}-\frac{e^{-\frac{3 c p}{2}+p \sin (\phi )+i p \cos (\phi )-i p}}{e^{c | p| }+1}-\frac{e^{-i p \Lambda_h}}{e^{-c|p|}+1}.
\end{align}
The total number of Bethe roots is then
\begin{equation}
    M=2+\int\sigma(\Lambda)\mathrm{d}\Lambda=\frac{1}{2} \left(N^e+1\right).
\end{equation}
Thus, the state has spin $S=0$, and the energy is $E_u$ because the energy of the string solution and the higher order string solution exactly cancel.

\subsection{Unscreened phase}

Notice that the impurity string solution still exists in this region. However, as shown in Eq.\eqref{istengus}, the energy of the impurity string solution vanishes. 
The energy of the string solution when $3c<2\sin(\phi)$  is calculated as
\begin{align}
    \Delta E^{\mathrm{ist}}&=D\int \Delta\sigma^{\mathrm{ist}}(\Lambda)[\Theta(2\Lambda-2)-\pi]\mathrm{d}\Lambda+D\left(\Theta\left(2 \Lambda_{stp}-2\right)- \pi\right)\nonumber\\
    &=2 i D \tanh ^{-1}\left(\frac{c-2 i e^{-i \phi }}{c}\right)-2 i D \tanh ^{-1}\left(\frac{c}{c-2 i e^{-i \phi }}\right)-\pi  D=0 .
\end{align}

Thus, for $3c<2\sin(\phi)$
\begin{equation}
    \Delta E^{\mathrm{ist}}=0.
    \label{istengus}
\end{equation}
Thus, in this phase, the impurity is always unscreened. 

The ground state is obtained by adding a hole to the continuous root distribution, which has energy
\begin{equation}
    E_{u}=-\frac{\pi}{2L}(N^e)^2-i D \log \left(\frac{\Gamma \left(\frac{i e^{-i \phi }}{2 c}\right) \Gamma \left(1+\frac{i e^{-i \phi }}{2 c}\right)}{\Gamma \left(\frac{c+e^{-i \phi } i}{2 c}\right)^2}\right)+2 D\left(\tan ^{-1} e^{(\pi / c)\left(\Lambda_{1}^{h}-1\right)}\right).
\end{equation}

Moreover, one can also add a hole and the impurity string solution to get a state $\ket{U}_s$, which is degenerate to the state $\ket{U}$ as the impurity string solution has vanishing energy. 

The root distribution in the $\ket{U}_s$ state is obtained by adding the discrete impurity string root string to the continuous root distribution. This makes the root distribution
\begin{eqnarray}
    \sigma_{imp}(\Lambda)=\frac{\frac{2}{-3 c+2 i (\Lambda -1)+2 \sin (\phi )+2 i \cos (\phi )}+\frac{\psi ^{(0)}\left(\frac{-c+2 i (\Lambda -1)+2 i \cos (\phi )+2 \sin (\phi )}{4 c}\right)-\psi ^{(0)}\left(\frac{c+2 i (\Lambda -1)+2 i \cos (\phi )+2 \sin (\phi )}{4 c}\right)}{c}}{2 \pi }.
\end{eqnarray}
Hence, the density of state for $\ket{U}_s$ state is
\begin{eqnarray}
    \rho(\mathcal{E})=\frac{\frac{2 \pi }{3 \pi -2 i \log \left(\frac{\mathcal E}{T_0}\right)}+\psi ^{(0)}\left(\frac{i \log \left(\frac{\mathcal E}{T_0}\right)}{2 \pi }-\frac{1}{4}\right)-\psi ^{(0)}\left(\frac{i \log \left(\frac{\mathcal E}{T_0}\right)}{2 \pi }-\frac{3}{4}\right)}{2 \pi ^2 \mathcal E}.
\end{eqnarray}
\end{appendix}

\end{document}